%% file: main.tex
\documentclass[a4paper]{jpconf}
\usepackage{graphicx}
\usepackage{tabularx}
\usepackage{xcolor}
\usepackage{lineno}

\usepackage[flushleft]{threeparttable}

\usepackage[square,sort&compress,numbers]{natbib}
\begin{document}
\title{Generative Adversarial Networks for the fast simulation of the Time Projection Chamber responses at the MPD detector}

\author{A~Maevskiy$^1$, F~Ratnikov$^{1,2}$, A~Zinchenko$^3$, V~Riabov$^4$, A~Sukhorosov$^1$ and D~Evdokimov$^1$}

\address{
$^1$ HSE University, 20 Myasnitskaya Ulitsa, Moscow, Russia)\\
$^2$ Yandex School of Data Analysis, 11-2 Timura Frunze Street, Moscow, Russia\\
$^3$ Joint Institute for Nuclear Research, 6 Joliot-Curie St, Dubna, Moscow Oblast, Russia\\
$^4$ Petersburg Nuclear Physics Institute, 1, mkr. Orlova roshcha, Gatchina, Leningradskaya Oblast, Russia
}

\ead{artem.maevskiy@cern.ch}

\begin{abstract}
The detailed detector simulation models are vital for the successful operation of modern high-energy physics experiments. In most cases, such detailed models require a significant amount of computing resources to run. Often this may not be afforded and less resource-intensive approaches are desired. In this work, we demonstrate the applicability of Generative Adversarial Networks (GAN) as the basis for such fast-simulation models for the case of the Time Projection Chamber (TPC) at the MPD detector at the NICA accelerator complex. Our prototype GAN-based model of TPC works more than an order of magnitude faster compared to the detailed simulation without any noticeable drop in the quality of the high-level reconstruction characteristics for the generated data. Approaches with direct and indirect quality metrics optimization are compared.
\end{abstract}

\section{Introduction}

Simulation of particle detectors is inevitable in the High Energy Physics (HEP) experiments. For a typical HEP data analysis, the limited size of simulated data samples often contributes directly to the uncertainty in the final result. Since the number of simulated events that one can afford to produce is constrained by the computational efficiency of the simulation algorithms, faster algorithms are always desired~\cite{HEPSoftwareFoundation:2017ggl}.

Computational efficiency of the detailed simulation is often limited by the fine granularity of the physics simulation steps being performed. Therefore, a speed-up may be achieved by aggregating a sequence of such steps with a single estimate of the probability distribution for the last step output parameters, conditioned by the first step inputs. An important requirement for such a probability distribution estimate is that it should allow for efficient sampling. Generative Adversarial Networks (GANs)~\cite{Goodfellow:2014upx} are a good candidate for such a parametric estimate since they only require a forward pass through a neural network to generate new samples. In this work, we demonstrate an application of GANs for building a fast-simulation model of the Time Projection Chamber (TPC) detector at the MPD experiment at the NICA accelerator complex~\cite{Maevskiy:2020ank}.

\section{TPC fast simulation approach and validation}
\label{sec:ourApproach}
The TPC detector is a gas-filled cylindrical volume with a uniform electric field parallel to its main axis. It performs charged particle tracking by measuring the 2D projection of the ionization centers locations, as the electrons drift through the gas volume and reach the sensitive pads at the end-caps of the chamber, the 3rd coordinate being reconstructed from the duration of the drift.

\begin{figure}[h]
\begin{minipage}{0.42\textwidth}
\centering
\includegraphics[height=14.6pc]{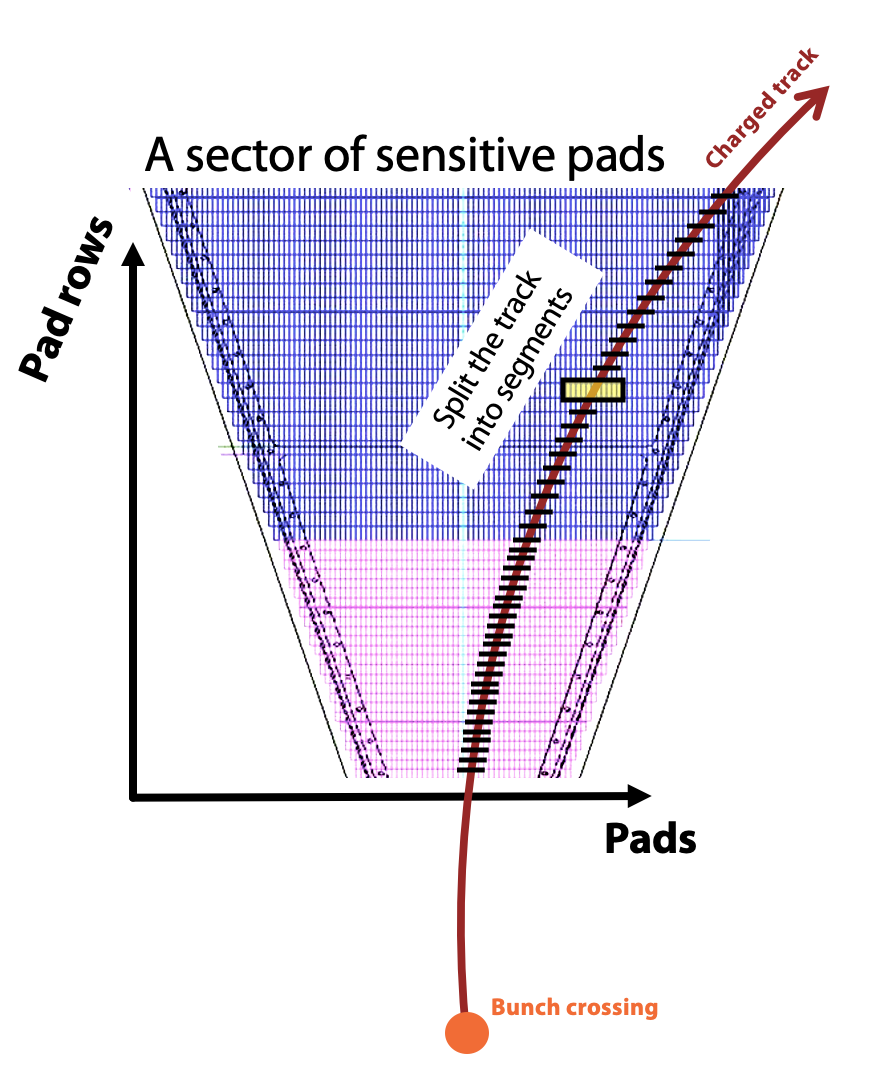}
\caption{\label{fig:approach}The transverse projection of a track on top of a sector of sensitive pads of the detector. The track is split into segments contributing to each of the rows of sensitive pads.}
\end{minipage}\hspace{0.04\textwidth}%
\begin{minipage}{0.52\textwidth}
\centering
\includegraphics[height=14.6pc]{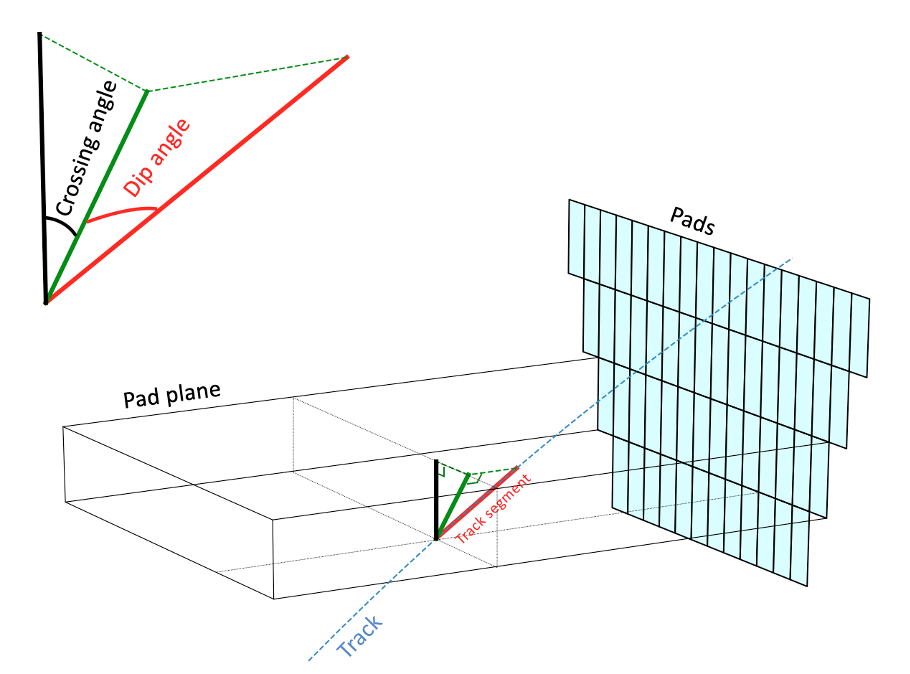}
\caption{\label{fig:angles}A track passing through the volume of the detector. An intersection of this track with a selected \emph{pad plane} forms a \emph{track segment} and is highlighted in the picture. The angles that define its direction are marked.}
\end{minipage} 
\end{figure}

The main steps that we aim to accelerate the simulation of are the ionization process, the electron drift and the electronics response~\cite{Maevskiy:2020ank}. We achieve this goal by training a GAN to reproduce the distribution of the read-out signals, as modeled by the detailed simulation procedure, conditioned on the parameters of the track segment contributing to the given response, as shown in Fig.~\ref{fig:approach}. For each track segment, we simulate a rectangular matrix of sensitive pad responses for a single row of pads in multiple discrete time intervals. Overall, we generate these responses, each in 8 pads for 16 time intervals, which makes a total of 128 numbers per segment. The track segments are parameterized by three spatial coordinates and two angles, as shown in Fig.~\ref{fig:angles}.

\begin{figure}[ht]
\begin{minipage}{0.65\textwidth}
\centering
\input{pre-eval-plots-matrix}
\caption{\label{fig:lowleveleval}Examples of low-level evaluation plots for our model. Shown are profiles for one of the six components of the lower-dimensional signal representation, compared between GAN and the detailed simulation as the function of the three (out of five) parameters defining the track segment. For more details, see~\cite{Maevskiy:2020ank}.}
\end{minipage}\hspace{0.025\textwidth}%
\begin{minipage}{0.32\textwidth}
\centering
\includegraphics[width=0.95\textwidth,trim=0 0 0 35,clip]{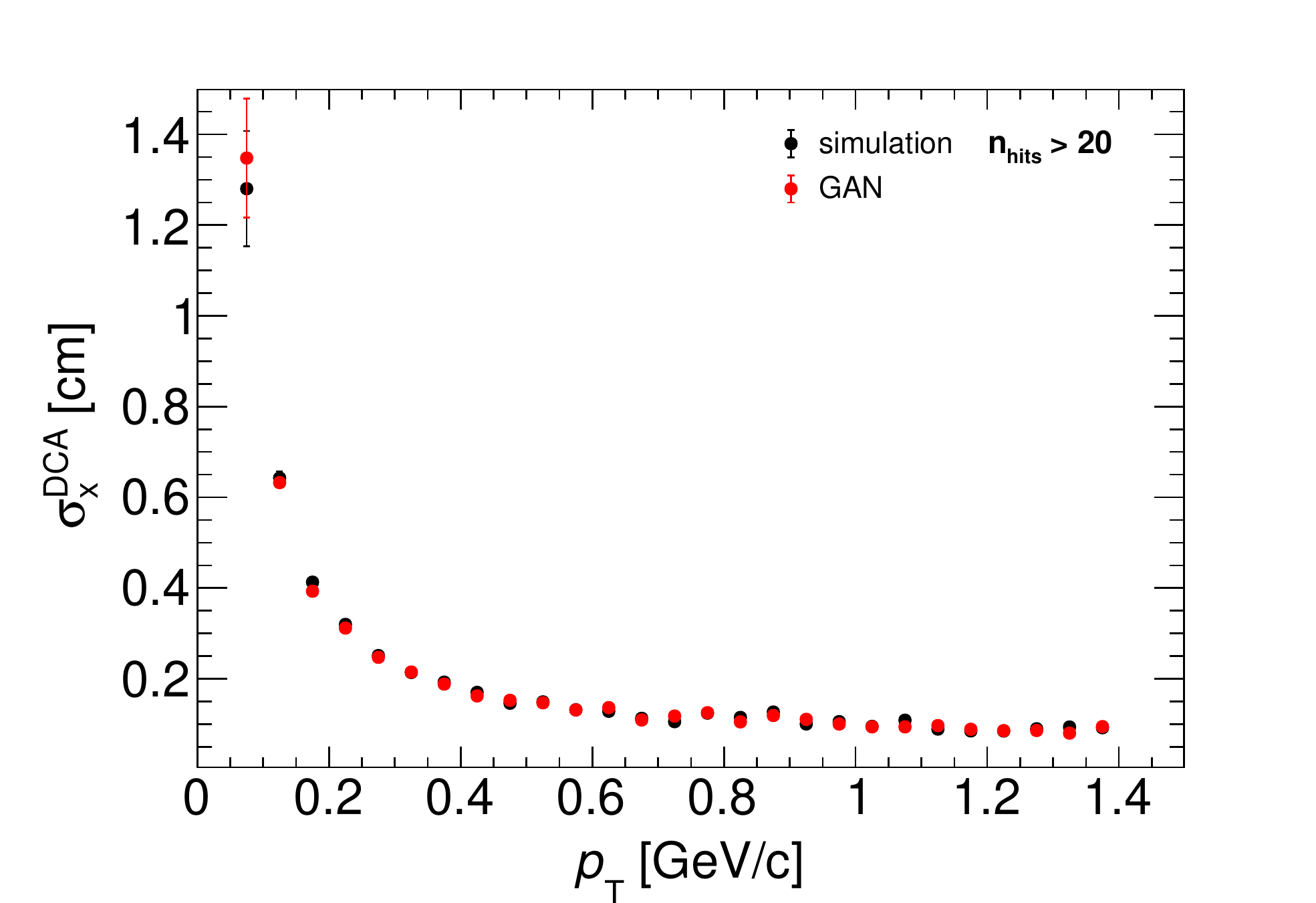}
\caption{\label{fig:resolution}An example of a high-level validation plot. Shown is the track distance of closest approach resolution as a function of the transverse momentum.}
\end{minipage} 
\end{figure}

We validate our model at two different levels:
\begin{itemize}
\item[-] The \emph{low-level evaluation} involves direct comparison of the GAN-generated detector response distribution with the left-out data sample of the training set.
\item[-] For the \emph{high-level evaluation} we compare the reconstructed track characteristics between the simulation configurations when our GAN and the detailed simulation models are used, respectively.
\end{itemize}
For the low-level evaluation, we transform the 128-dimensional responses into lower-dimensional representation which is then compared between the fast and detailed simulation models, coordinate-wise, as the function of the input parameters. The lower-dimensional representation is obtained by calculating the coordinates of the centers of mass for the obtained responses, along with their widths, skew and the integrated amplitude~--- six numbers per response in total. These quantities are chosen from our expectations on the factors that affect the TPC tracking characteristics the most. E.g., the center of mass locations translate directly to the reconstructed hit coordinates and therefore should affect the coordinate resolution of the detector. Similarly, response widths should relate to the two-track resolution. We characterize the distributions of these numbers with their means and standard deviations in 1D bins of the input variables. Some examples of the low-level evaluation plots for such quantities are shown in Fig.~\ref{fig:lowleveleval}. Also, for model selection and to control the convergence of the training process, we calculate a $\chi^2$-like quantity by accumulating the squared discrepancies between the mean $\pm$ one standard deviation values of such distributions in each 1D bin.

The high-level evaluation is done by comparing the reconstructed track characteristics between the fast and detailed TPC simulation configurations. Fig.~\ref{fig:resolution} shows an example of such a comparison plot. Overall, both validation strategies demonstrate good quality of the proposed fast simulation model. For a more detailed validation, see~\cite{Maevskiy:2020ank}.

Our model speeds up the TPC simulation by at least a factor of 12, with some room for further inference time optimization, as described in the following section.

\section{Inference time optimization}

Once the model is successfully trained and validated, it is crucial to deploy it in the most computationally efficient way. Having developed the baseline model with good enough quality and inference time, as described in the previous section, we then explore the ways of further decreasing the inference time without a loss in quality. We study the model performance in terms of time required for predicting a batch of detector responses on a single CPU thread. Such configuration corresponds to the expected target execution platform. For a single-point quality estimate, we use the $\chi^2$-like quantity defined in the previous section.

\begin{figure}[ht]
\begin{minipage}{0.48\textwidth}
\centering
\includegraphics[width=0.98\textwidth,trim=30 40 40 40,clip]{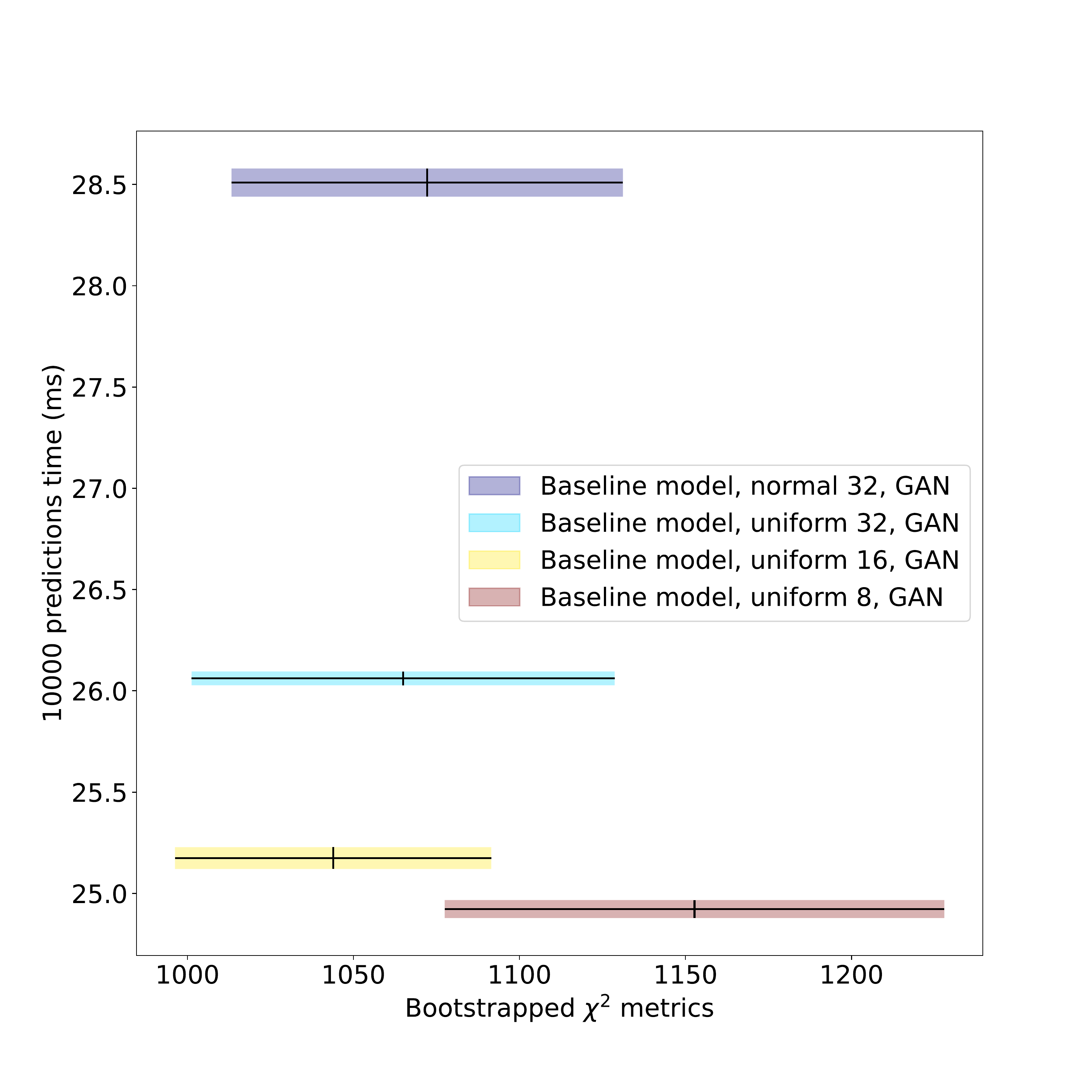}
\caption{Inference time versus quality for various latent space configurations.}
\label{fig:latent_space}
\end{minipage}\hspace{0.025\textwidth}%
\begin{minipage}{0.48\textwidth}
\centering
\includegraphics[width=0.98\textwidth,trim=40 40 30 40,clip]{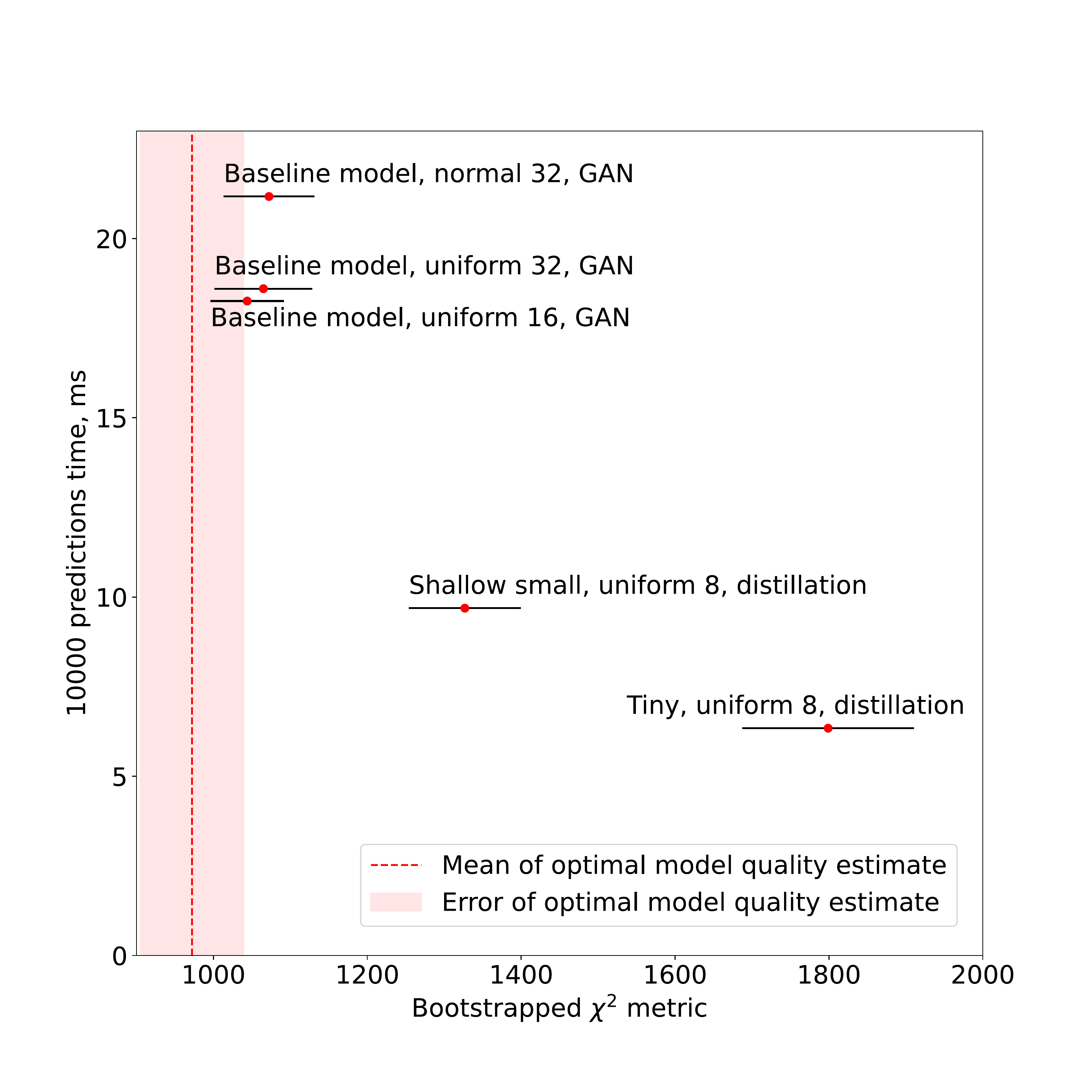}
\caption{Inference time versus quality for different network architectures.}
\label{fig:comparison}
\end{minipage} 
\end{figure}

The generator of the baseline model is a fully-connected network consisting of four hidden layers of sizes ranging from 32 to 64 and a 32-dimensional normally distributed latent variable. By profiling the inference of this generator we observe that, for such a small network, the latent space random number generation takes a notable fraction of the inference time. This means that we can optimize the inference time by choosing a simpler latent representation. Fig.~\ref{fig:latent_space} shows that models based on uniformly distributed latent variables allow for faster inference without any loss in quality, unless we decrease the size of this representation down to 8.

Optimizing the architecture for a network that is already small is tricky. Further reducing the size of the model results in poor generation quality and unstable training within the GAN framework. We observe that the best quality of such smaller models can be achieved by training them with the knowledge distillation technique~\cite{hinton2015distilling}, the baseline generator being used as the teacher. The resulting experimental quality-speed trade-off curve is shown in Fig.~\ref{fig:comparison}. It can be seen that, while latent space optimizations do allow for faster models of the same quality, further simplifying the network architecture results in a quality drop.

\section{Direct low-level quality metric optimization}

One may wonder: why bother generating the high-dimensional detector response if its major characteristics can be defined with just six numbers that we use in the low-level evaluation, as described in Sec.~\ref{sec:ourApproach}? Indeed, we may learn just the distribution of those six parameters, and then deterministically transform them into the higher-dimensional signals. In such a scenario, our generative model aims to directly reproduce the low-level quality metrics, as they get explicitly incorporated in the optimized objective function. In this section, we explore this low-level optimization approach and highlight some pitfalls that make it not as straightforward as one may expect.

\begin{figure}[ht]
\begin{minipage}{0.65\textwidth}
\centering
\input{pre-eval-DE}
\caption{\label{fig:lowlevelevalDE} Examples of low-level evaluation plots for a model generating the low-level metrics directly. Shown are the profiles for one of the low-level characteristics.}
\end{minipage}\hspace{0.025\textwidth}%
\begin{minipage}{0.32\textwidth}
\centering
\includegraphics[width=0.95\textwidth,trim=20 7 30 20,clip]{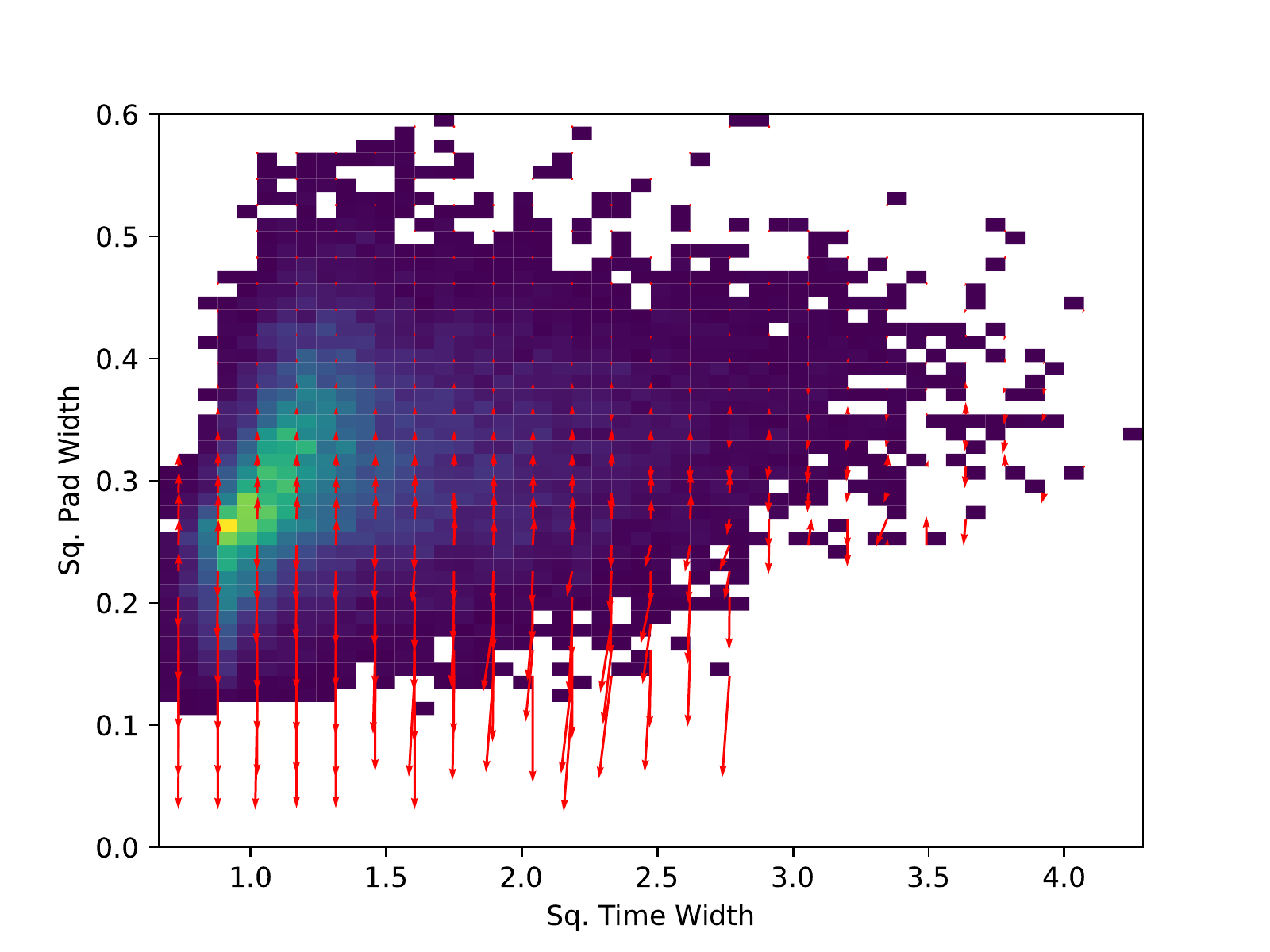}
\caption{\label{fig:momentsDistortion}The squared pad and time widths distribution, with the distortion directions (see text).
}
\end{minipage} 
\end{figure}

First of all, when trying to learn the 6-dimensional distribution of the signal characteristics, we find that it is much harder to reach a reasonable quality level. The quality metrics for our best such model are shown in Fig.~\ref{fig:lowlevelevalDE}. One can clearly see that there are noticeable discrepancies between the GAN-generated and training samples. We acknowledge that this result may be a consequence of an imperfect hyper-parameter search and therefore continue to look for a more optimal configuration.

Another issue with this approach is that it is not obvious how to transform the six low-level characteristics back into the 128-dimensional (8-by-16) response matrix. Since the chosen characteristics can be thought of as the 0-th (integral), 1-st (center of mass location) and 2-nd (widths and skew) moments of a 2D distribution, it may be natural to describe the inverse transformation by a probability density function (normalized to the required integral) of a 2D distribution that does not have higher moments~--- i.e., the Gaussian distribution, parameterized by the six mentioned quantities. The problem is that the target 2D distribution should be quantized onto a rectangular 8-by-16 grid, while this operation does not preserve the moments. Fig.~\ref{fig:momentsDistortion} shows the distribution of the two widths for the signals in the training dataset, superimposed with the arrows denoting the distortion for given pairs of widths, if we were to convert the signals to quantized Gaussians. Each arrow has the length of the average distortion magnitude and points towards the most popular distortion direction, within each 2D bin. One can clearly see that the level of distortion is too high to be ignored. A solution that we are looking into is to incorporate the distorting operation between the generator and the discriminator networks during training, such that the generator learns to compensate for it automatically.

\section{Summary}

In this work, we present a fast simulation approach for the TPC detector using GANs. We demonstrate that our model accelerates the detailed simulation by at least an order of magnitude, and it is capable of producing detector responses that look authentic in both low- and high-level validation procedures. The trade-off between the inference time and generation quality is further explored by optimizing the model architecture. An alternative solution with direct low-level quality metric optimization is also outlined.

\section*{Acknowledgments}

The work is partially supported by the HSE University Project Group "Fast simulation of the TPC tracker at the MPD experiment with deep learning" (order no. 2.3-05/310821-3).

This research was supported in part through computational resources of HPC facilities at NRU HSE.

\bibliography{refs}

\end{document}

%% file: pre-eval-plots-matrix.tex
\begingroup
\tiny
\setlength{\tabcolsep}{0pt}
\begin{tabular}{>{\raggedleft}m{.035\textwidth}>{\centering}m{.3\textwidth}>{\centering}m{.3\textwidth}>{\centering\arraybackslash}m{.3\textwidth}}
{\rotatebox[origin=c]{90}{\parbox{.29\textwidth}{\centering Sq. Pad Width}}} &
\raisebox{-.5\height}{\includegraphics[width=.29\textwidth,trim=0 10 35 30,clip]{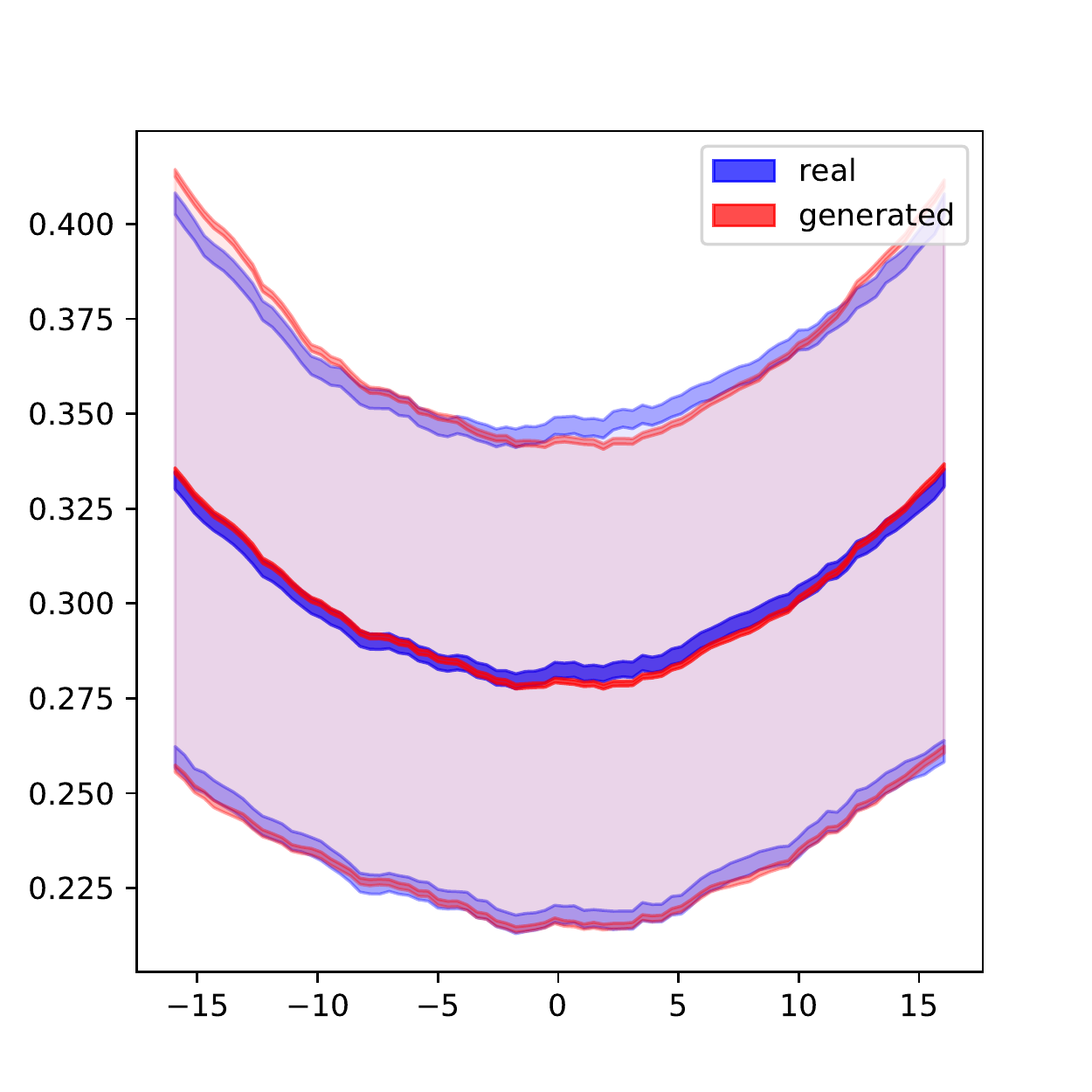}} &
\raisebox{-.5\height}{\includegraphics[width=.29\textwidth,trim=0 10 35 30,clip]{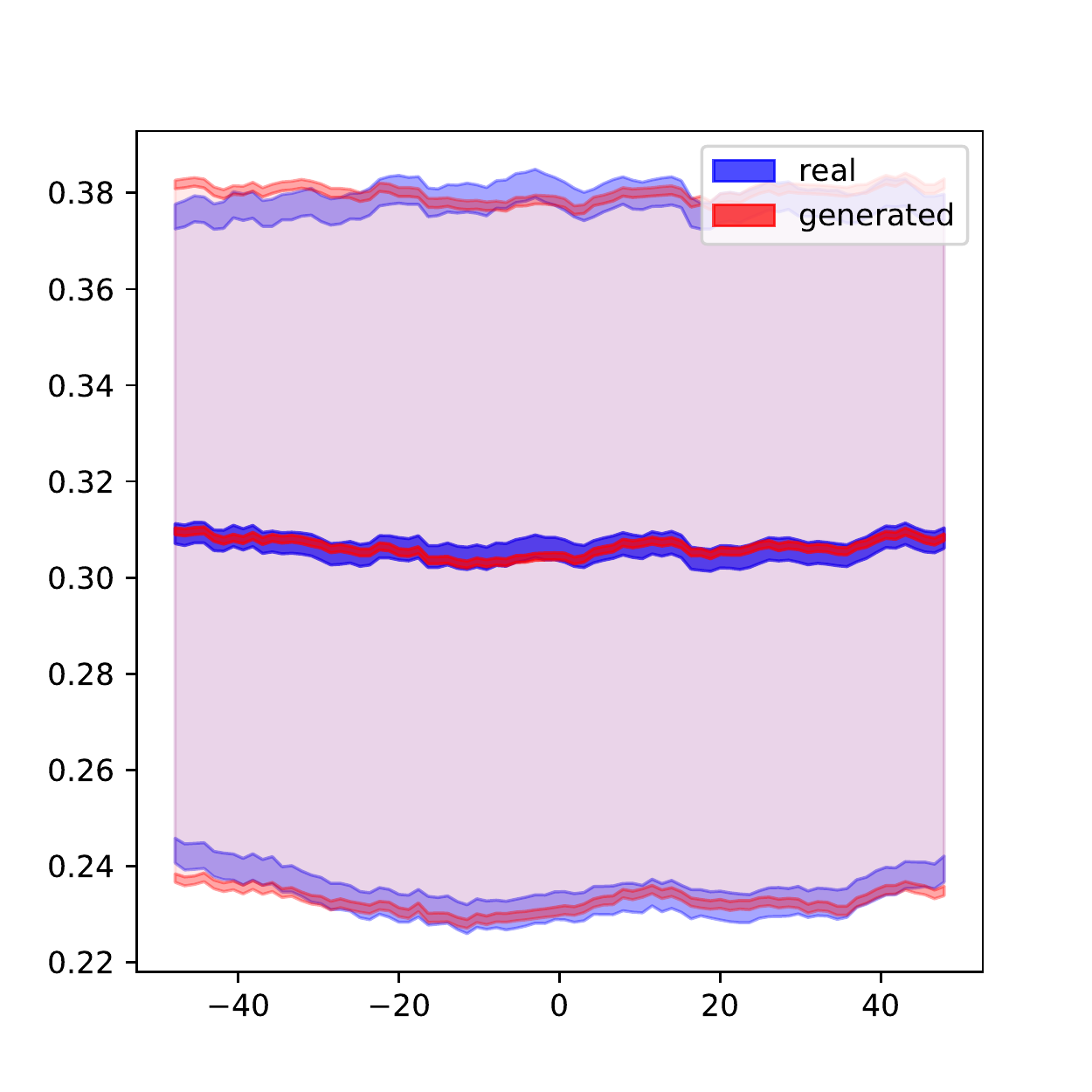}} &
\raisebox{-.5\height}{\includegraphics[width=.29\textwidth,trim=0 10 35 30,clip]{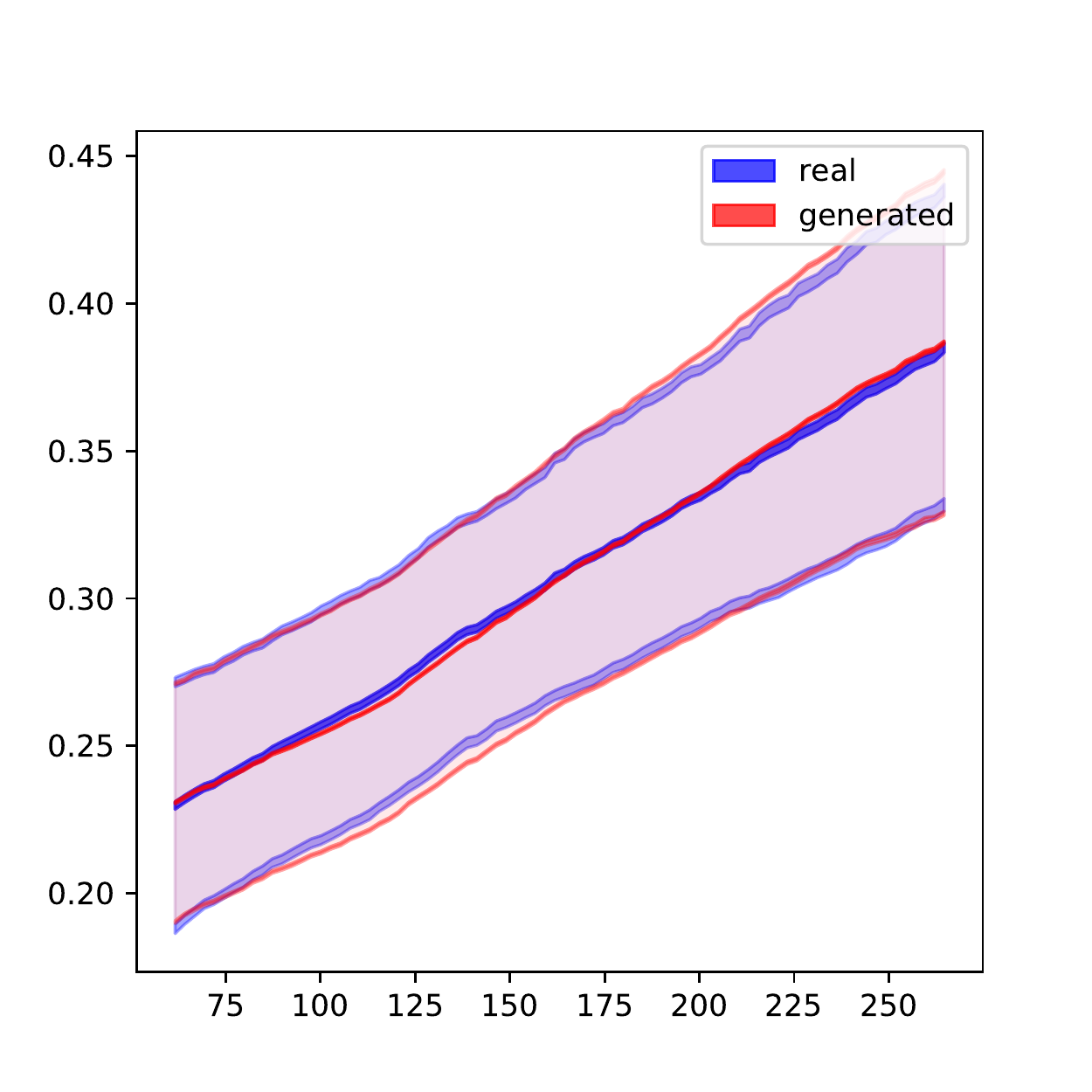}}\\
~ &
\hspace{10pt}\parbox{.15\textwidth}{\centering Crossing angle [deg]} &
\hspace{10pt}\parbox{.15\textwidth}{\centering Dip angle [deg]} &
\hspace{10pt}\parbox{.15\textwidth}{\centering Drift length [time bins]}
\end{tabular}
\endgroup

%% file: pre-eval-DE.tex
\begingroup
\tiny
\setlength{\tabcolsep}{0pt}
\begin{tabular}{>{\raggedleft}m{.035\textwidth}>{\centering}m{.3\textwidth}>{\centering}m{.3\textwidth}>{\centering\arraybackslash}m{.3\textwidth}}
{\rotatebox[origin=c]{90}{\parbox{.29\textwidth}{\centering Sq. Pad Width}}} &
\raisebox{-.5\height}{\includegraphics[width=.29\textwidth,trim=0 10 35 30,clip]{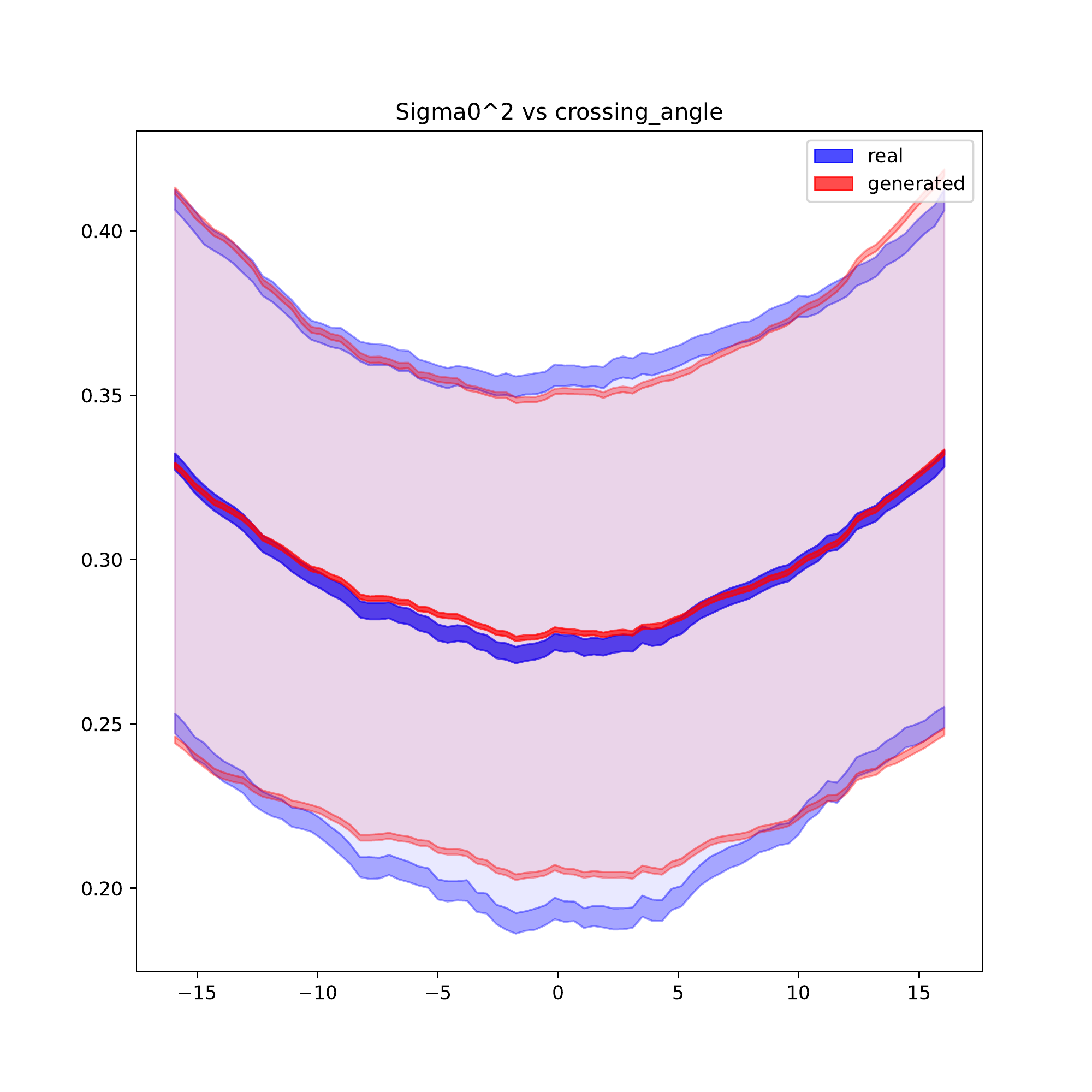}} &
\raisebox{-.5\height}{\includegraphics[width=.29\textwidth,trim=0 10 35 30,clip]{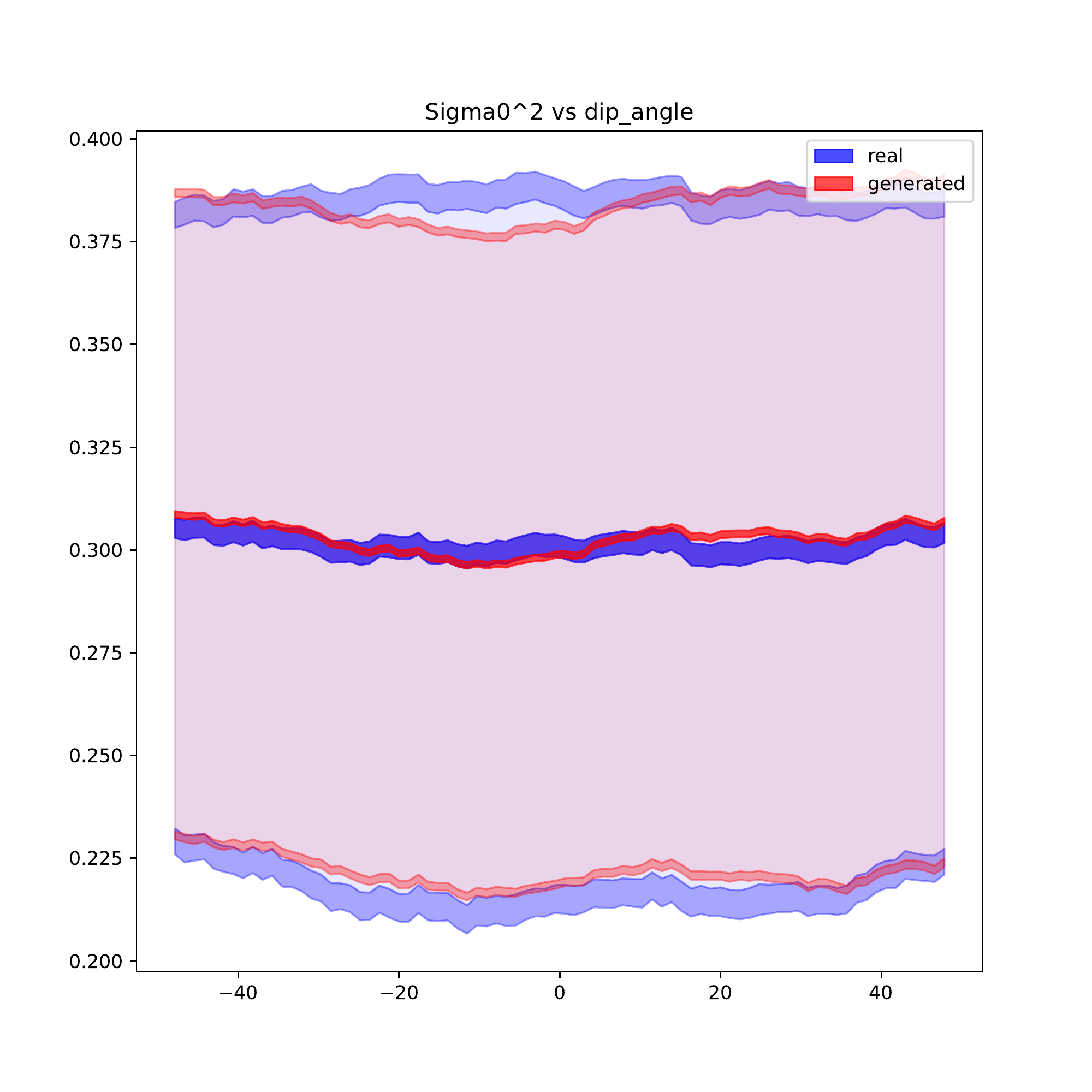}} &
\raisebox{-.5\height}{\includegraphics[width=.29\textwidth,trim=0 10 35 30,clip]{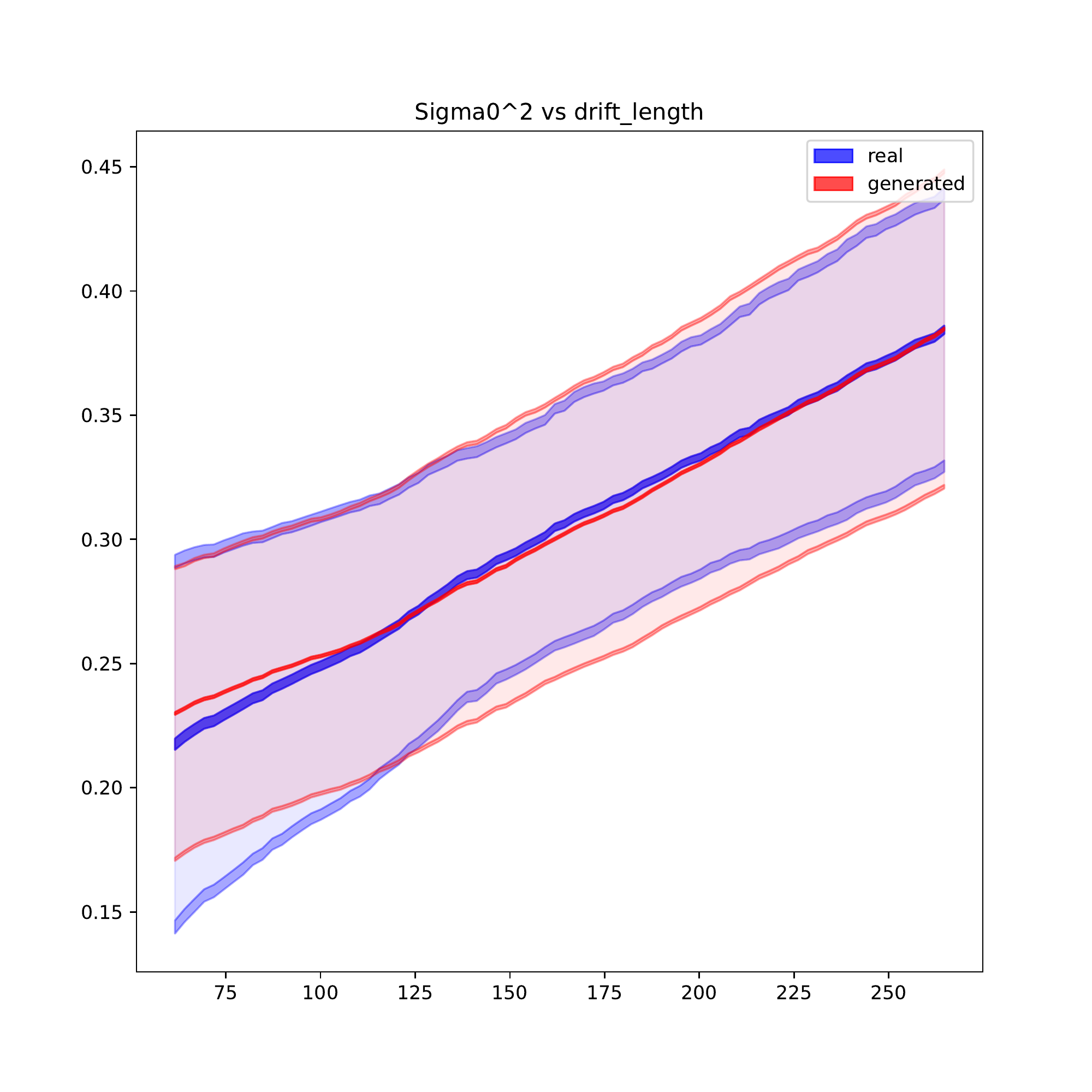}}\\ 
~ &
\hspace{10pt}\parbox{.15\textwidth}{\centering Crossing angle [deg]} &
\hspace{10pt}\parbox{.15\textwidth}{\centering Dip angle [deg]} &
\hspace{10pt}\parbox{.15\textwidth}{\centering Drift length [time bins]}
\end{tabular}
\endgroup